\documentclass[pra,twocolumn,showpacs]{revtex4-1}

\usepackage[english]{babel}
\usepackage{amssymb}
\usepackage{amsfonts}
\usepackage{amsmath}
\usepackage{graphicx}
\usepackage{color}
\usepackage{subfig}
\usepackage{psfrag}
\usepackage{epstopdf}
\usepackage{placeins}

\DeclareMathOperator{\sech}{sech}

\newcommand{\myfigfont}{\footnotesize}
\newcommand{\intpar}[1]{\! \! \mathrm{d}#1 \ }
\newcommand{\intpartreeNS}[1]{ \mathrm{d}^{3}#1 \ }
\newcommand{\intparNS}[1]{ \mathrm{d}#1 \ }
\newcommand{\Exp}[1]{\mathrm{e}^{#1}}
\newcommand{\abs}[1]{\left| #1 \right|}

\begin{document}

\title{Scattering of matter wave solitons on localized potentials}

\author{Sidse Damgaard Hansen, Nicolai Nygaard, and Klaus M\o lmer}
\affiliation{
Lundbeck Foundation Theoretical Center for Quantum System Research, Department of
Physics and Astronomy, University of Aarhus, DK-8000 Aarhus C, Denmark.}

\date{\today}

\begin{abstract}
We present numerical and analytical results for the reflection and transmission properties of matter wave solitons impinging on localized scattering potentials in one spatial dimension. Our mean field analysis identifies regimes where the solitons behave more like waves or more like particles as a result of the interplay between the dispersive wave propagation and the attractive interactions between the atoms. For a bright soliton propagating together with a dark soliton void in a two-species Bose-Einstein condensate of atoms with repulsive interactions, we find different reflection and transmission properties of the dark and the bright components.

\end{abstract}

 %Solitons in BEC, tunneling in BEC. Non-linear dynamics of solitons. Matter waves.

\maketitle

\section{Introduction}

Solitons are exceptionally stable wave phenomena that appear in a variety of physical systems. They propagate without spreading due to a fine balance between the dispersion and the non-linearity of the physical system. After their discovery as stable waves in shallow waters by Russell in 1834, solitons have been found in very different non-linear systems. As an example, both bright optical solitons which are peaks in the intensity and dark solitons which are intensity minima have found applications in optical transmission through fibres. In this article we consider soliton waves in Bose-Einstein condensates of neutral atoms.

The condensation of a cold gas of atoms permits a description of the many-body problem by a single condensate wave function order parameter, which is in turn described by the non-linear Gross-Pitaevskii wave equation \cite{bog}. The non-linearity in Bose-Einstein condensates arises from the repulsive or attractive interactions of the particles in the condensate. In equivalence with non-linear optics, dark and bright solitons are, respectively, localized dips and peaks in the Bose-Einstein condensate density. 

The dynamics of solitons in inhomogeneous condensates with slowly varying potentials have been investigated and is adequately described by a particle-like behavior \cite{solitontrain,solitontrain2,4,5,BandADarkSoliton, burger, reinhardt, denschlag}. In this article, we shall further elaborate the analysis of matter wave solitons in 1D-condensates, investigating their scattering properties on localized potentials.

In Sec. II, we introduce the Gross-Pitaevskii equation and a convenient mapping to dimensionless coordinates that will be used throughout the paper. In Sec. III, we introduce the bright, the dark, and the composite dark-bright solitons, and we briefly review their properties. In Sec. IV, we show that scattering of bright solitons on a low and wide potential barrier can be well understood as the motion of a classical particle governed by an effective Lagrangian, extracted from the quantum problem, while the scattering on high and narrow barriers can be understood from the single particle quantum problem. In Sec. V we present results for the scattering of a dark-bright soliton on a potential felt only by the bright soliton atoms, and we show to what extent their scattering is transferred to the dark soliton. Sec. VI concludes the paper.

\section{Mean field theory of interacting atoms}
We treat the Bose-Einstein condensate of atoms at zero temperature in a Hartree mean field theory, which neglects particles outside the condensate and treats short range interactions in the gas by an interaction potential which is proportional to the mean atomic density.
This yields the 1D timedependent Gross-Pitaesvkii equation (GPE)
\begin{align}
i\hbar\frac{\partial \psi(x,t)}{\partial t}&= \bigl(-\frac{\hbar^2}{2m}\frac{\partial ^2}{\partial x^2} +V(x,t)\label{NGPEt}\\\nonumber 
&+g\abs{ \psi(x,t)} ^2\bigr)\psi(x,t),
\end{align}
where the order parameter, the macroscopic wave function, is normalized to the total number of atoms
\begin{equation}
 \int \intpartreeNS{r}\abs{\psi(r)} ^2 =N
\end{equation}
so that $|\psi(x,t)|^2$ represents the particle density. The 1D interaction strength $g$ is a known function of the 3D low energy scattering properties and the transverse confinement of the atoms \cite{bog}.\par

In an attractive BEC with $g<0$, (meta-)stable bright soliton solutions with a finite number of particles exist, while in a repulsive BEC with $g>0$, dark solitons may exist in the form of holes in the background density.\par 

For numerical purposes it is advantageous to rewrite the GPE in a dimensionless form, and for a homogeneous condensate with repulsive interactions, the linear atomic density, $n_0$ and the healing length $\xi=\sqrt{\frac{\hbar^2}{mgn_0}}$ provide natural scale factors. We thus introduce dimensionless quantities for length, energy and time, labelled by a tilde
\begin{align}
&x=\tilde{x}\xi , \qquad E=\tilde{E}\frac{\hbar ^2}{m\xi ^2}=\tilde{E}n_0 g,\qquad t=\tilde{t}\frac{m\xi ^2}{\hbar}.
\label{xEt}
\end{align}
and obtain the equation

\begin{equation}
i\frac{\partial \tilde{\psi}}{\partial \tilde{t}}=-\frac{1}{2}\frac{\partial ^2 \tilde{\psi}}{\partial \tilde{x}^2}+\abs{\tilde{\psi}} ^2\tilde{\psi} + \tilde{V}\tilde{\psi}.
\label{dimensionlessGPErep}
\end{equation}%

where the dimensionless wave function

\begin{equation}
\tilde{\psi}=\frac{\psi}{\sqrt{n_0}},
\label{tildepsi}
\end{equation}

is normalized to unit bulk density $\abs{\tilde{\psi}}^2 =1$ in the homogeneous part of the condensate.

A condensate with attractive interactions has no natural length scale like the healing length, but for convenience in the following, we introduce an arbitrary length unit $l$, which may be chosen to fit a relevant experimental length scale, and we set
\begin{equation}
x=\tilde{x}l, \qquad E=\tilde{E}\frac{\hbar^2}{ml^2}, \qquad t=\tilde{t}\frac{ml^2}{\hbar}.
\end{equation}
Inserting this in the GPE, Eq. (\ref{NGPEt}), and introducing the rescaled wave function, $\abs{\tilde{\psi}}^2=\abs{ g}{\frac{ml^2}{\hbar^2}}\abs{\psi}^2$, leads to the dimensionless GPE,
\begin{equation}
i\frac{\partial \tilde{\psi}}{\partial \tilde{t}}=-\frac{1}{2}\frac{\partial^2 \tilde{\psi}}{\partial \tilde{x}^2}-\abs{\tilde{\psi}}^2\tilde{\psi} +\tilde{V}\tilde{\psi}.
\end{equation}
The rescaled number of particles in the condensate is
\begin{equation}
\tilde{N}=\int \intparNS{\tilde{x}}\abs{\tilde{\psi}} ^2 = N\frac{ml\abs{ g}}{\hbar ^2}.
\end{equation}
Thus for given $N$, the choice of $\tilde{N}$ determines the length scale $l$ and sets the dimensions of the problem.

Due to the presence of the non-linear term it is in general necessary to solve the Gross-Pitaesvkii equation by numerical means. The numerical results presented in the following are based on propagation of the one-dimensional, dimensionless and non-homogeneous GPE using the Crank-Nicolson method. For brevity we will omit the tilde above dimensionless quantities throughout.

\section{Solitons}
\label{solitons}
In this section we recall the soliton solutions of the homogeneous Gross-Piaevskii equation.

\subsection{Dark solitons}
Although the scattering properties of dark solitons are not discussed in this paper we will first introduce the dark soliton. This will serve as a background for the later discussion of the dark-bright soliton.\par
Dark solitons are formed in repulsive BECs, governed by the homogeneous GPE, Eq. (\ref{dimensionlessGPErep}), and fulfilling the boundary condition $\lim_{x\rightarrow \pm \infty}\abs{\psi(x)}^2=n_0$, where $n_0$ is the bulk value of the condensate density and also the dimensionless chemical potential. A dark soliton soliton moving with velocity $v$ is then described by 
\begin{align}
&\psi_\mathrm{D}(x,t)=\sqrt{n_0}\Biggl[i\frac{v}{c}+\sqrt{1-\frac{v^2}{c^2}}\tanh\left(\frac{x-vt}{\xi_v}\right)\Biggr]\Exp{-i\mu t/\hbar},
\label{dark}
\end{align}
where $\xi_v=\xi/\sqrt{1-\frac{v^2}{c^2}}$, with $\xi$ the healing length and $c=\sqrt{\frac{n_0 g}{m}}$ the speed of sound, both unity in our dimensionless units \cite{bog}. 
The dark soliton has a density dip around $x=vt$, and we will characterize the size of the soliton by the number of atoms missing, $N_\mathrm{D}$, compared to the homogeneous case,
\begin{align}
N_\mathrm{D}&=\int \intparNS{x} \left(n_0-\abs{\psi_\mathrm{D}} ^2\right)=2n_0\xi\sqrt{1-\frac{v^2}{c^2}}.
\label{singelN_D}
\end{align}
The density depression moves in the opposite direction of the atoms in the condensate, whose motion is given by the gradient of the phase of the condensate wave function. The energy of the dark soliton is given by the difference between the energy of the condensate containing a dark soliton and the energy of the homogeneous condensate,
\begin{align}
E_\mathrm{D}&=\int \intparNS{x} \Bigl(\frac{\hbar ^2}{2m}\abs{\frac{\partial\psi_\mathrm{D}}{\partial x}} ^2+\frac{g}{2}\left( \abs{\psi_\mathrm{D}} ^2 -n_0\Bigr) ^2
\right)\\\nonumber
&=\frac{4}{3}n_0\hbar c \left(1-\frac{v^2}{c^2}\right)^{3/2}.
\end{align}
The soliton energy decreases with increasing soliton velocity, and for $\frac{v}{c}\ll 1$ one finds
\begin{equation}
E_\mathrm{D}\simeq \frac{4}{3}n_0\hbar c -\frac{2n_0\hbar}{c}v^2,
\end{equation}
corresponding to an effective negative mass of $-4n_0\frac{\hbar}{c}$.

\subsection{Bright solitons}

If the interactions in the BEC are attractive ($g<0$) the homogeneous GPE supports bright soliton solutions,
\begin{align}
\psi_\mathrm{B}(x,t)=&\sqrt{\frac{2\mu}{g}}\sech\Bigl(\frac{\sqrt{2m\abs{\mu}}}{\hbar}(x-vt)\Bigr)\Exp{i(mvx-\omega t)/\hbar},
\label{brightunits}
\end{align}
where $\omega =\frac{1}{2}mv^2 +\mu$, $\mu$ being the chemical potential \cite{bog}. The bright soliton propagates with velocity $v$, and contrary to the dark soliton, its shape does not depend on its velocity, but only on the number of atoms in the condensate.

In the dimensionless coordinates, the soliton solution, Eq. (\ref{brightunits}), is
\begin{align}
\psi_\mathrm{B}(x,t)=A\sech \left[ B(x-vt)\right] \Exp{i\left(vx-\omega t\right)},
\label{solution}
\end{align}
where $A=B=N/2$, the chemical potential of the condensate is $\mu = -\frac{B^2}{2}$, and $\omega =\frac{v^2}{2}-\frac{B^2}{2}$.

The energy of the bright soliton, in the absence of an external potential is given by three contributions, \cite{6}:
\begin{equation}
E=E_\mathrm{kin}^\mathrm{d}+E_\mathrm{kin}^v+E_\mathrm{int}.
\end{equation}
The first part is the quantum pressure contribution to the kinetic energy 
\begin{equation}
E_\mathrm{kin}^\mathrm{d}=\frac{1}{2}\int \intparNS{x} \abs{ \frac{\partial \abs{\psi}}{\partial x}} ^2 = \frac{N^3}{24}.
\label{Edkin}
\end{equation}

This is present even when the soliton is stationary, while the contribution to kinetic energy from the phase gradient
\begin{align}
E_\mathrm{kin}^v&=\frac{1}{2}\int \intparNS{x} \abs{ \abs{\psi}\frac{\partial \Exp{i(vx-\omega t)}}{\partial x}} ^2=\frac{1}{2}Nv^2,
\end{align}
disappears for a stationary soliton.\par
Finally, the interactions between the atoms contribute 
\begin{equation}
E_\mathrm{int}=-\frac{1}{2}\int \intparNS{x} \abs{\psi} ^4 =-\frac{N^3}{12}
\label{Eint}
\end{equation}
to the energy of the soliton.\par 

In total the energy of the bright soliton is

\begin{equation}
 E_\mathrm{B}=\frac{1}{2}Nv^2-\frac{N^3}{24}.
\end{equation} 

\subsection{Dark-Bright solitons}
\label{darkbrightsolitons}
A dark-bright soliton is a composite object in a two-component condensates consisting of a bright soliton BEC and a density depression in a repulsive BEC. Here, we consider the case where all interactions, internal and between the two components, are repulsive with equal strengths, and in the rescaled coordinates, the coupled Gross-Pitaevskii Equations of the system read

\begin{gather}
\begin{aligned}
&i\frac{\partial \psi_\mathrm{1}(x,t)}{\partial t}=\Bigl(-\frac{1}{2}\frac{\partial ^2}{\partial x^2}+\abs{\psi_\mathrm{1}(x,t)} ^2\\%\nonumber
&\phantom{{}i\frac{\partial \psi_\mathrm{1}(x,t)}{\partial t}=\Bigl(} +\abs{\psi_\mathrm{2}(x,t)} ^2\Bigr)\psi_\mathrm{1}(x,t),\\%\nonumber
&i\frac{\partial \psi_\mathrm{2}(x,t)}{\partial t}=\Bigl(-\frac{1}{2}\frac{\partial ^2}{\partial x^2}+\abs{\psi_\mathrm{2}(x,t)} ^2\\%\nonumber
&\phantom{{}i\frac{\partial \psi_\mathrm{2}(x,t)}{\partial t}=\Bigl(} +\abs{\psi_\mathrm{1}(x,t)} ^2\Bigr)\psi_\mathrm{2}(x,t),
\end{aligned}
\label{twocomponentBEC}
\end{gather}
This non-linear, coupled set of differential equations has a dark-bright soliton solution:
%
%\begin{gather}
%\begin{aligned}
\begin{align}
\nonumber &\psi_\mathrm{D}(x,t)=\Bigl(i\sqrt{\mu}\sin \alpha+\sqrt{\mu}\cos\alpha\tanh\left[\kappa (x-x_0)\right]\Bigr)\Exp{-i\mu t},\\\nonumber
&\psi_\mathrm{B}(x,t)=\sqrt{\frac{N_\mathrm{B}\kappa}{2}}\Exp{ix\kappa\tan\alpha}\sech\left[\kappa (x-x_0)\right]\cdot \\
&\phantom{{}\psi_\mathrm{B}(x,t)=}\Exp{-i\left(\frac{\kappa ^2\tan^2\alpha}{2}-\frac{\kappa^2}{2}+\mu\right)t},
\label{darkbrightsolitonsolution}
\end{align}%\end{aligned}

where
\begin{align}
\kappa^2+\frac{\kappa N_\mathrm{B}}{2}=\mu\cos^2\alpha,
\label{kappaligning}
\end{align}

and
\begin{align}
x_0=vt,
\label{center}
\end{align}
with $v=\kappa\tan \alpha$ \cite{Becker, BogA}.

The bright soliton can only exists in the repulsive condensate because it is trapped inside the dark soliton.
The dark-bright soliton is wider than the single dark soliton, and the larger the number of particles in the bright component, the wider it is due to the repulsive interaction between the two components.

The energy of the two-component condensate can be separated into three parts,
\begin{equation}
E_\mathrm{DB}=E_\mathrm{D} +E_\mathrm{B} +E_\mathrm{int},
\end{equation}
where the energy of the dark soliton $E_\mathrm{D}$ is defined by subtracting the energy of the dark background condensate,
\begin{gather}
\begin{aligned}
E_\mathrm{D}&=\int \intparNS{x} \left( \frac{1}{2}\abs{\frac{\partial\psi_\mathrm{D}}{\partial x}} ^2
+\frac{1}{2}\left( \abs{\psi_\mathrm{D}} ^2 -\mu\right) ^2\right)\\
&=\frac{2}{3}\kappa\mu\cos^2\alpha +\frac{2}{3}\frac{\mu^2\cos^4\alpha}{\kappa},
\end{aligned}
\end{gather}
$E_\mathrm{B}$ is the energy of the bright soliton
\begin{gather}
\begin{aligned}
E_\mathrm{B}&=\int\intparNS{x} \left(\frac{1}{2}\abs{\frac{\partial\psi_\mathrm{B}}{\partial x}} ^2 +\frac{1}{2}\abs{\psi_\mathrm{B}} ^4\right)\\
&= \frac{1}{2}N_\mathrm{B}v^2 +\frac{1}{6}N_\mathrm{B}\kappa\left(N_\mathrm{B} +\kappa\right) ,
\end{aligned}
\end{gather}
and $E_\mathrm{int}$ is the energy caused by the interaction between the two components
\begin{align}
E_\mathrm{int}& =\int \intparNS{x}\abs{\psi_\mathrm{B}} ^2\abs{\psi_\mathrm{D}} ^2\\\nonumber
&=N_\mathrm{B}\mu\sin^2\alpha +\frac{1}{3}N_\mathrm{B}\mu\cos^2\alpha .
\end{align}
The total vector soliton energy can then be written as \citep{BogA}
\begin{equation}
E_\mathrm{DB}=\frac{4}{3}\kappa ^3 +\frac{1}{2}N_\mathrm{B}\kappa^2 + \frac{1}{2}N_\mathrm{B}v^2 +\mu N_\mathrm{B}.
\end{equation}

\section{Scattering of Bright solitons}
\label{Scatteringofbrightsolitons}
Bright solitons are localized objects and in weakly varying potentials where the non-linearity is important, they largely behave as classical particles \cite{shoulder,1,2,3,11,Ching-Hao}, whereas in strongly varying potentials we expect wave diffraction phenomena to enter \cite{Ching-Hao,2,3,11,lee,6,7,weiss,holmer,delta,lang,Parker}. In this Section, we shall investigate the transition between particle and wave behavior of bright solitons scattering on simple square potentials in one dimension.

\subsection{Weak Barriers}
\label{weakbarriers}
As mentioned in the previous section, the chemical potential of a bright soliton is given by $\mu=-\frac{N^2}{8}$, and when the potential is weak compared to the chemical potential, $N^2\gg V_\mathrm{ex}$, we expect the strong interaction potential to hold the atoms together throughout the scattering process.\par

\begin{figure*}[htbps]
\myfigfont
\centering
\subfloat[\label{v=0.24}]{\includegraphics[scale=1]{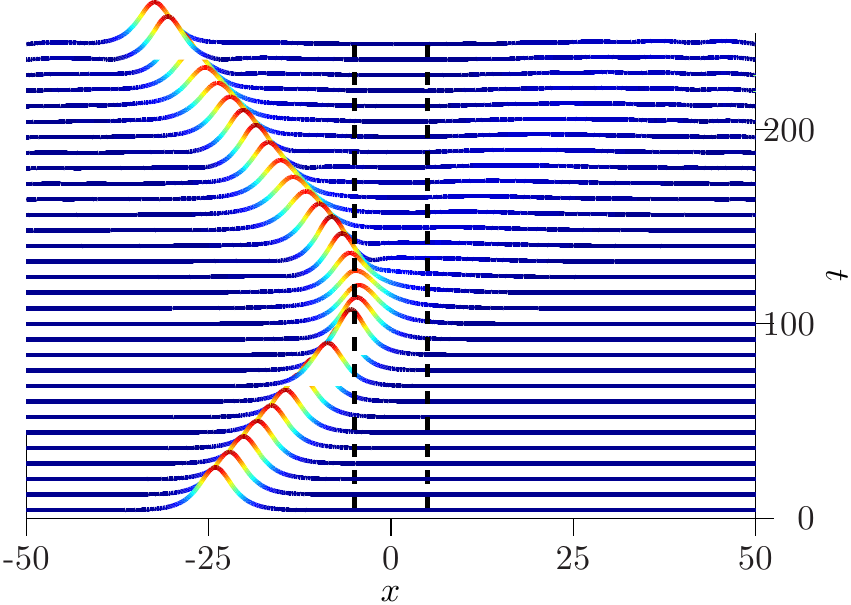}}
\qquad
\subfloat[\label{v=0.29}]{\includegraphics[scale=1]{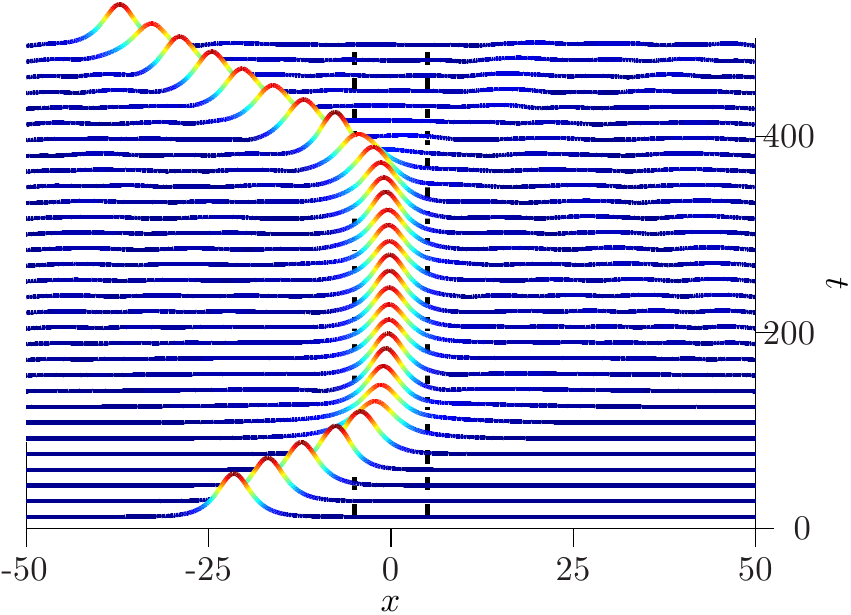}}
\qquad
\subfloat[\label{v=0.3}]{\includegraphics[scale=1]{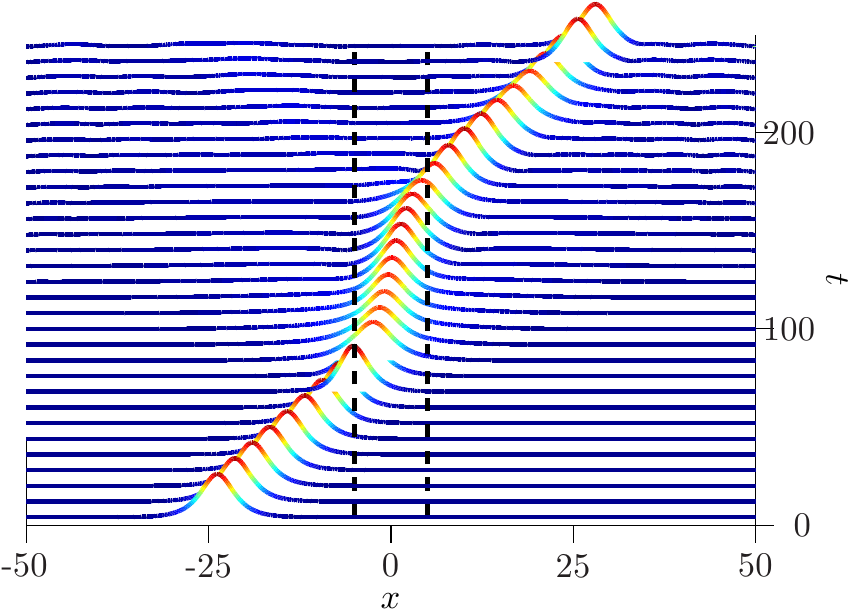}}
\qquad
\subfloat[\label{v=0.4}]{\includegraphics[scale=1]{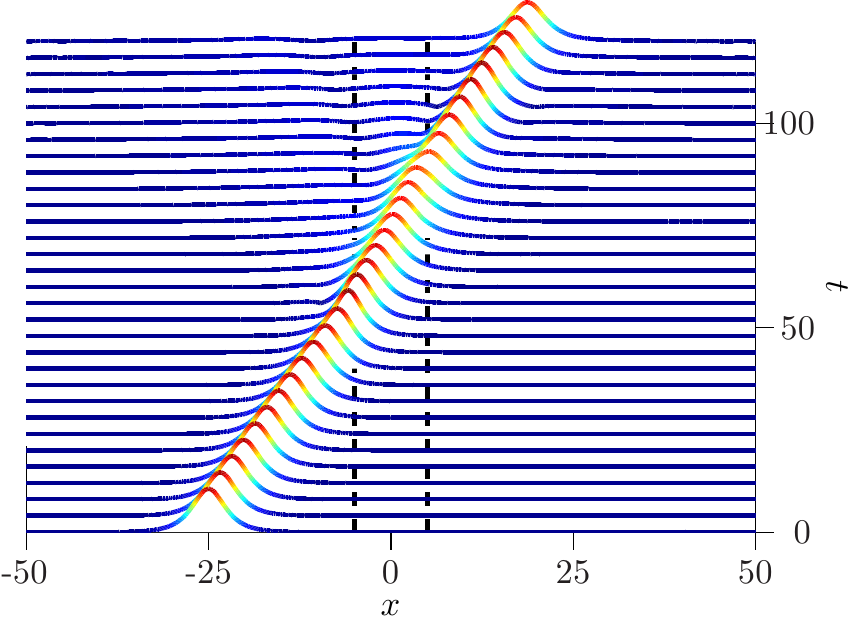}}
\caption{(color online) Time evolution of the condensate density. The external potential, indicated by the dashed lines, is a square barrier with height $V_0=0.04$ and width $b=10$ and the incoming soliton has $N=1$. The threshold velocity, Eq. (\ref{v_crit}), is $v_\mathrm{t}\approx 0.289$. In \protect\subref{v=0.24} the incoming soliton velocity is $v=0.24$, which is well below the threshold velocity, in \protect\subref{v=0.29} the velocity is $v=0.29$, which is just about the threshold velocity and in \protect\subref{v=0.3} and \protect\subref{v=0.4} the velocities are $v=0.3$ and $v=0.4$ respectively, which are above the threshold velocity.}
\label{classical}
\end{figure*}

In figure \ref{classical} the time evolution of the Gross-Pitaevskii wave function is shown for soliton wave packets, incident on a weak square barrier with four different incident velocities. The calculations show that the soliton keeps its sech-shape and is either reflected or transmitted by the barrier like a classical particle.\par
In Figs. \ref{v=0.29} and \ref{v=0.3} the soliton is slowed down dramatically and dwells on the potential for a rather long time, before it is eventually reflected or transmitted regaining its initial shape and speed. 

\begin{figure*}[]
\myfigfont
%\centering
\subfloat[\label{b=10}]{\includegraphics[scale=0.99]{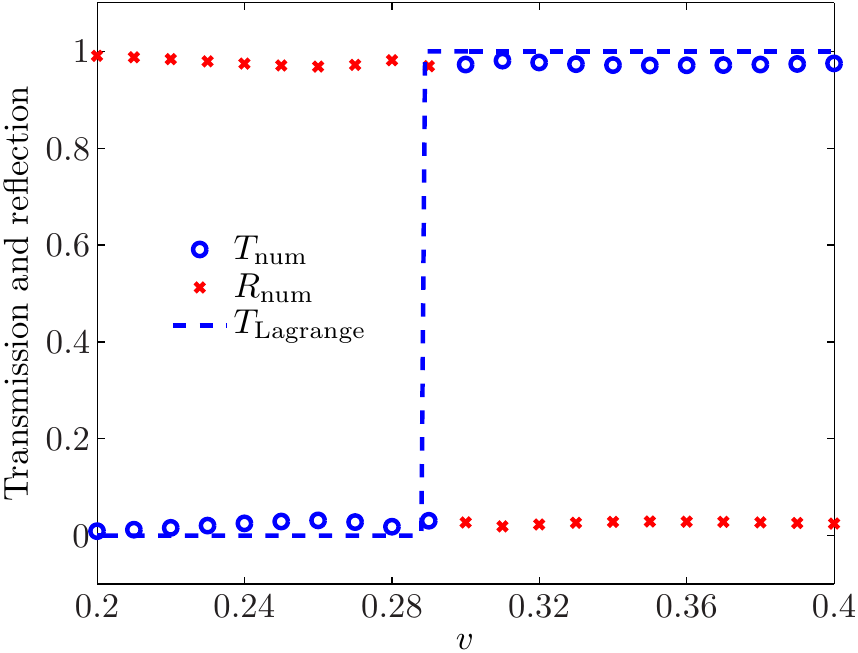}}
\qquad
\subfloat[\label{b=2}]{\includegraphics[scale=0.99]{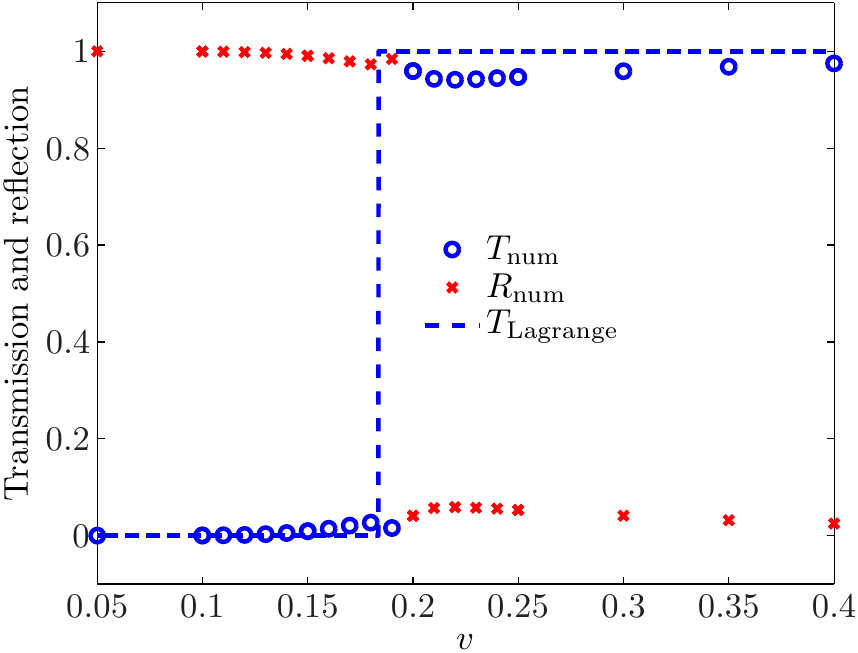}}
\caption{(Color online) Transmission and reflection coefficients of a soliton scattering on a weak square barrier for various soliton velocities found by a numerical solution of the GPE (symbols) and by the Lagrangian method (dashed lines). The soliton has $N=1$ and the barrier height is $V_0=0.04$. In \protect\subref{b=10} the barrier has width $b=10$, and is thus wide compared to the soliton width $B^{-1}=2$. In \protect\subref{b=2} the barrier has width $b=2$ which is comparable to the soliton width.}
\label{klassisk}
\end{figure*}

We define reflection and transmission coefficients as the weight of the wave function on the left and right hand side of the barrier
\begin{gather}
\begin{aligned}
&R=\frac{1}{N}\int_{-L/2}^0 \intpar{x}\abs{ \psi} ^2
\\
&T=\frac{1}{N}\int_0^{L/2} \intpar{x}\abs{ \psi} ^2,
\end{aligned}
\label{TandRnum}
\end{gather}
where $L$ is the length of the box used for our calculations, and the barrier is centred at $x=0$. In Fig. \ref{klassisk} transmission and reflection coefficients are shown for two different barrier widths. The figure confirms that the solitons are almost exclusively reflected or transmitted by the wide barrier, while the narrow barrier causes a minor splitting of the mean field wave packet.\par

To get a better understanding of the observed phenomena, we assume that the soliton keeps its sech-shape during the scattering process, a fair assumption as seen from the numerical calculations in Fig. \ref{classical}, 
\begin{equation}
\psi(x,t)=A\sech \left[ B(x-x_0)\right] \Exp{i\dot{x}_0\left(x-x_0\right)+i\phi},
\end{equation}

and we use the Lagrange approach presented in \cite{1,2,3} to identify equations of motion for the time dependent parameters $A$, $B$, $x_0$ and $\phi$. The effective Lagrangian of the system is
\begin{gather}
\begin{aligned}
\mathcal{L}&=\int_{-\infty}^\infty \intpar{x}\left(\frac{i}{2}\left(\frac{\partial \psi}{\partial t}\psi^\ast - \frac{\partial \psi^\ast}{\partial t}\psi\right)\right) -E(\psi)\\
&=\frac{1}{2}N\dot{x}_0^2-N\dot{\phi}+\frac{1}{6}NB(N-B)-V_{\mathrm{eff}}(x_0),
\end{aligned}
\label{lagrange}
\end{gather}
where $V_{\mathrm{eff}}(x_0)$ is
\begin{equation}
V_{\mathrm{eff}}(x_0)= \int_{-\infty}^\infty \intpar{x}V(x)\abs{\psi(x)}^2.
\label{Veffdef}
\end{equation}
We are now able to find the classical equations of motion for the soliton from the Euler-Lagrange equations
\begin{equation}
\frac{d}{d t}\frac{\partial \mathcal{L}}{\partial \dot{q}}=\frac{\partial \mathcal{L}}{\partial q}, \qquad q=B,x_0,\phi.
\label{euler}
\end{equation}
Eq.(\ref{euler}) with $q=\phi$ yields $\dot{N}=0$, i.e., the particle number is conserved.\par
Eq. (\ref{euler}) with $q=x_0$ yields the equation of motion for the location of the soliton
\begin{equation}
\ddot{x}_0=-\frac{1}{N}\frac{\partial V_{\mathrm{eff}}(x_0)}{\partial x_0}.
\end{equation}
This is the equation of motion of a classical particle with mass $N$ and center of mass $x_0$ moving in the effective potential, $V_\mathrm{eff}$, defined by Eq. (\ref{Veffdef}).\par
Finally Eq. (\ref{euler}) with $q=B$ yields 
\begin{equation}
N^2-2NB=6\frac{\partial V_{\mathrm{eff}}(x_0)}{\partial B}.
\label{Beq}
\end{equation}
describing a variation of the soliton width as it propagates within the potential barrier.\par
For a square barrier, i.e., an external potential of the form
\begin{equation}
V(x)=\begin{cases}
0 & \text{for $\abs{ x} >\frac{b}{2}$}\\
V_0 & \text{for $\abs{ x} <\frac{b}{2}$},
\end{cases}
\label{barrier}
\end{equation}
the effective potential becomes
\begin{align}
V_{\mathrm{eff}}(x_0)= \begin{cases}
0 & \text{for $\abs{ x} >\frac{b}{2}$}\\
\frac{V_0N}{2}\bigl(\tanh \left[B\left(\frac{b}{2}-x_0\right)\right]\\
+\tanh\left[B\left(\frac{b}{2}+x_0\right)\right]\bigr) & \text{for $\abs{ x} <\frac{b}{2}$}.
\end{cases}
\label{Veff}
\end{align}

The Hamiltonian of the motion is found from the Lagrangian:
\begin{equation}
H=N\dot{x}_0^2-\mathcal{L}=\frac{1}{2}N\dot{x}_0^2+\frac{1}{6}NB(B-N)+V_\mathrm{eff}(x_0),
\end{equation}
and the threshold incident energy the soliton needs to cross the barrier is given by 
\begin{equation}
\frac{1}{2}Nv_\mathrm{t}^2-\frac{N^3}{24}=\mathrm{max}_{\abs{ x_0} < \frac{b}{2}}\left\lbrace \frac{1}{6}NB(B-N)+V_\mathrm{eff}(x_0)\right\rbrace .
\label{v_crit}
\end{equation}
The corresponding threshold velocity of the bright soliton is indicated by the dashed lines in Fig. \ref{klassisk}, and reproduces the transition between complete reflection and transmission by the potential.\par
We also note that since the barrier is softened by the interparticle interactions, Eq. (\ref{Veff}), it makes sense that the soliton penetrates into the classical forbidden areas as seen in Fig. \ref{classical} \cite{shoulder}. 

\subsection{Strong Barriers}
\label{strongbarriers}
Scattering of bright solitons on strong barriers with $V_\mathrm{ex}\gg \mu$ has been studied in the limit of high soliton velocities \cite{2,holmer,6,delta}. In this regime the soliton wave function may coherently split into reflected and transmitted components, and the asymptotic behavior of the transmission and reflection as the soliton velocity increases is the same as for a single particle, described by the linear Schr\"odinger equation. Here, we shall consider scattering of bright solitons on strong potentials with special attention to the behavior at low soliton velocities.

\begin{figure*}[htbp]
\myfigfont
\centering
\begin{minipage}[b]{0.47\textwidth}
%\centering
\includegraphics[scale=0.96]{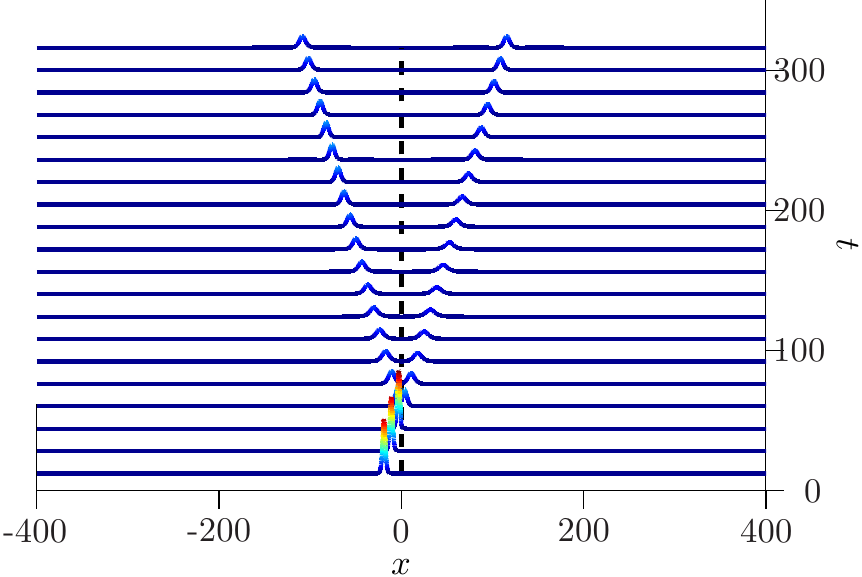}
\end{minipage}
\hfill
\begin{minipage}[b]{0.47\textwidth}
%\centering
\includegraphics[scale=0.96]{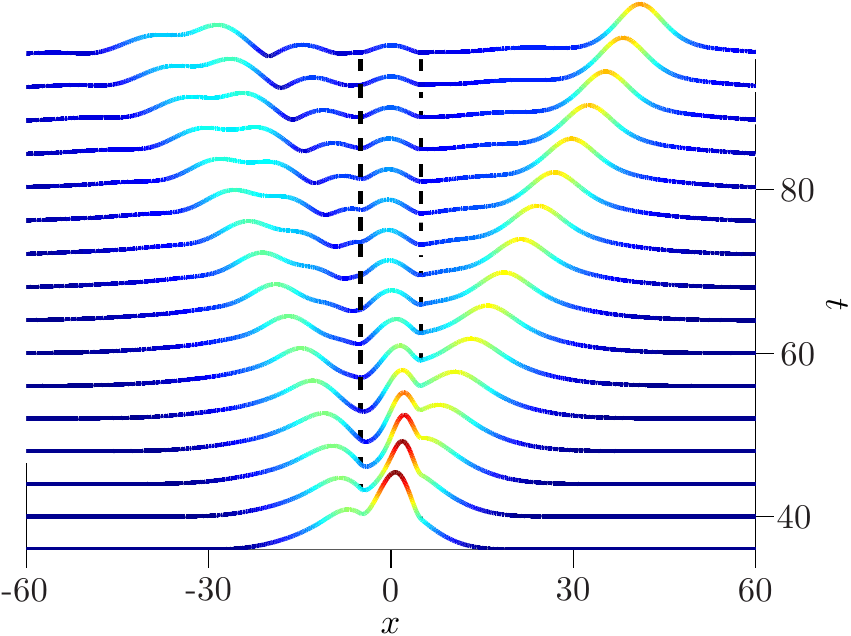}
\end{minipage}
\begin{minipage}[t]{0.47\textwidth}
\caption{(Color online) Time evolution of the condensate density. The external potential is a square barrier placed at $x=0$. The potential has height $V_0=0.8$ and width $b=0.5$, and the incoming soliton has $v=0.5$ and $N=1$.}
\label{waterfallV08b05v05}
\end{minipage}
\hfill
\begin{minipage}[t]{0.47\textwidth}
\caption{(Color online) Time evolution of the condensate wave function. The external potential is a square barrier placed at $x=0$. The potential has $V_0=0.2$ and $b=10$, and the incoming soliton velocity is $v=0.7$. Here we have zoomed in on the time evolution of the wave function $\abs{\psi (x)}$ (not the density), in order to see the details better. }
\label{zoom}
\end{minipage}
\end{figure*}

Fig. \ref{waterfallV08b05v05} shows the time evolution of a bright soliton incident on a rectangular potential barrier. We observe both a reflected and a transmitted component in the calculation. When the potential barrier is high enough to split the soliton, i.e. $V_0 \gg \mu$, and at the same time wide compared to the soliton width, a part of the soliton is seen to be transiently trapped inside the barrier, see Fig. \ref{zoom}. This trapping inside a potential barrier is due to the wave packet nature of the soliton. The trapped component oscillates and is gradually emitted out of the barrier region. This behavior is not seen when the barrier is narrow compared to the soliton width.

\begin{figure*}[]
\myfigfont
%\centering
\subfloat[\label{V=0.6}]{\includegraphics[scale=1]{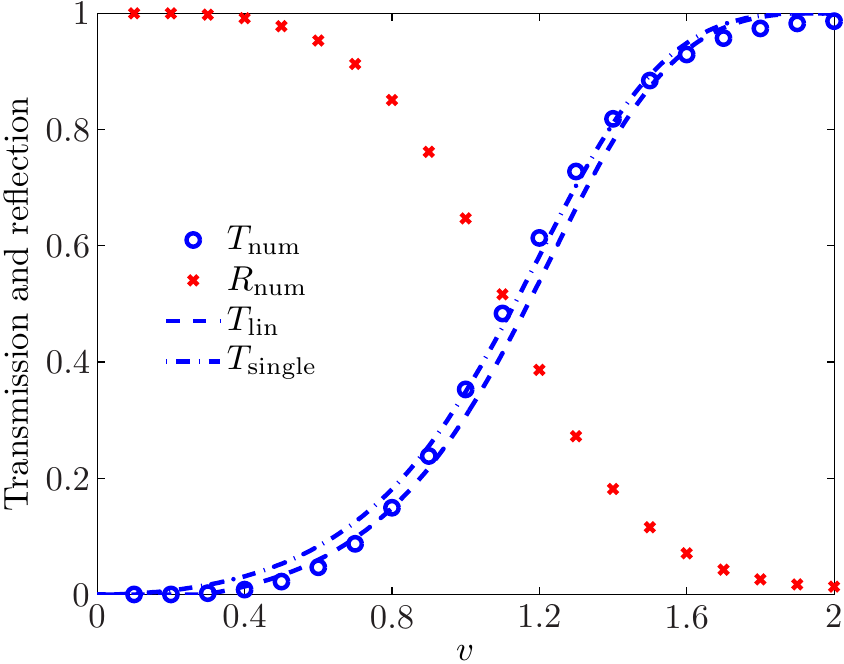}}
\qquad
\subfloat[\label{V=0.2}]{\includegraphics[scale=1]{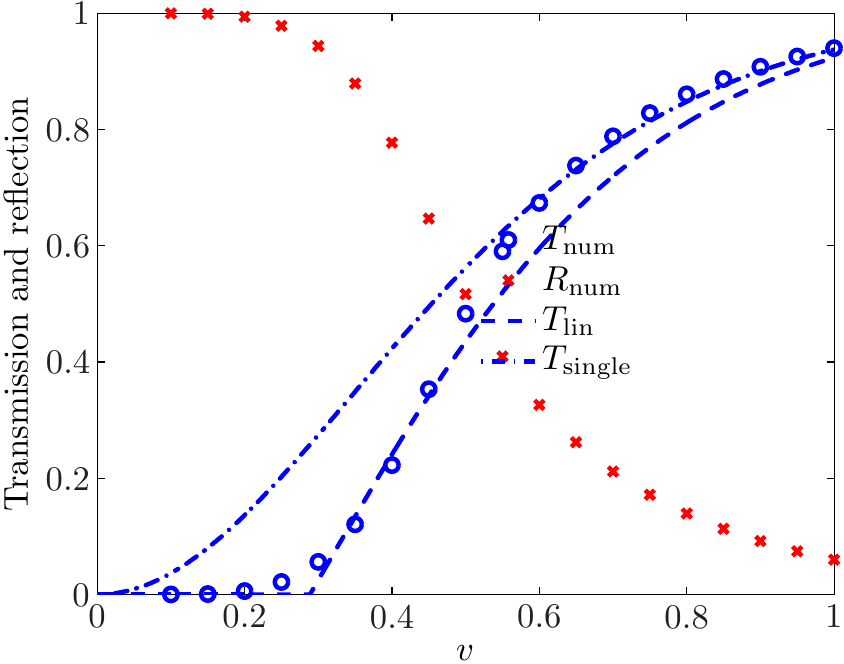}}
\qquad
\subfloat[\label{V=0.8}]{\includegraphics[scale=1]{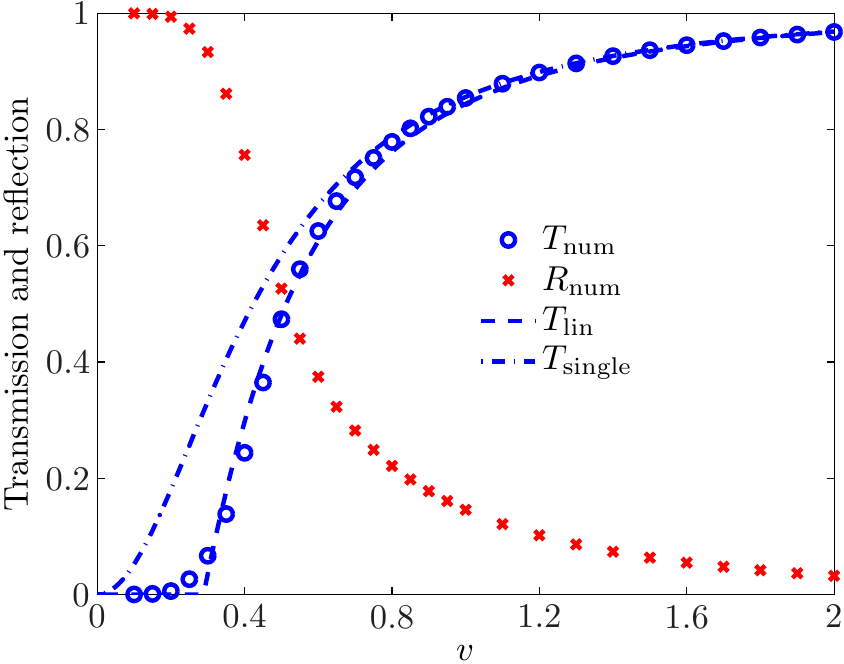}}
\qquad
\subfloat[\label{V=0.4}]{\includegraphics[scale=1]{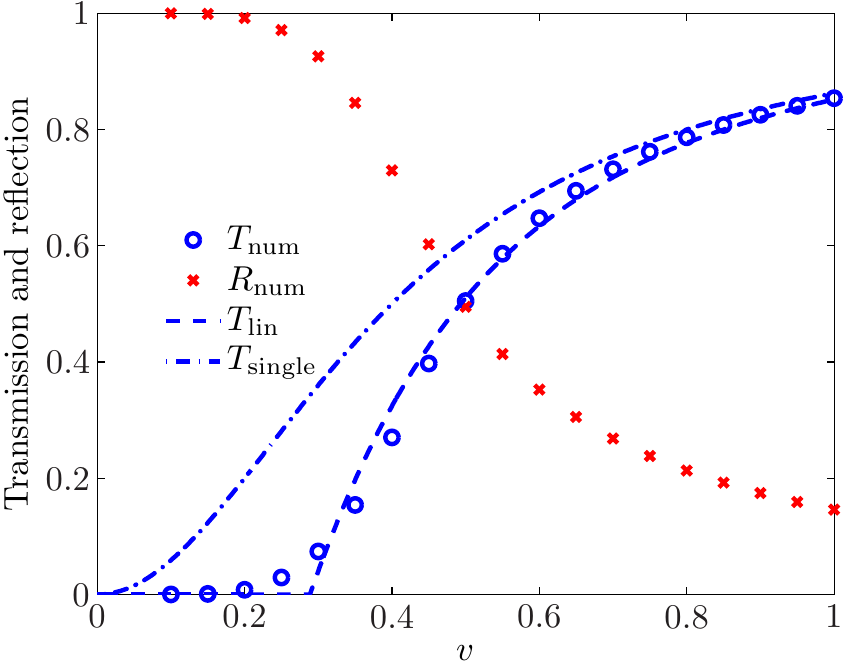}}
\caption{(Color online) Transmission and reflection coefficients of a soliton with $N=1$ on a strong barrier for various incoming soliton velocities. The coefficients has been found by solving the GPE numerically (symbols), by Eq. (\ref{Transmission}) with $E = \frac{1}{2}v^2$ (dashed-dotted lines), and from Eq. (\ref{Transmission}) with $E=\frac{1}{2}v^2-\frac{N^2}{24}$ (dashed lines), see the text. In \protect\subref{V=0.6} the barrier has strength $V_0=0.6$ and width $b=2$, in \protect\subref{V=0.2} the barrier has strength $V_0=0.2$ and strength $b=2$, and in \protect\subref{V=0.8} the barrier has $V_0=0.8$ and $b=0.5$, and thus the same area as in Fig. \protect\subref{V=0.2}. 
All barriers are higher than the chemical potential, $\abs{ \mu} =\frac{1}{8}$. 
In \protect\subref{V=0.4} the barrier is a delta barrier of strength $\alpha = 0.4$ which means that the barrier has the same area as the barriers in \ref{V=0.2} and \ref{V=0.8}. For the delta potential the transmission probability reads $T=\Theta (E)\cdot\frac{1}{1+a^2}, \qquad a=\frac{\alpha}{\sqrt{2E}}$.}
\label{kvant}
\end{figure*}

We have solved the GPE numerically for different cases, and Fig. \ref{kvant} summarizes our results (symbols) as functions of the incident velocity. When the external potential is strong compared to the chemical potential of the soliton, we expect the interaction term in the GPE to be less important, and the results in Fig. \ref{kvant} may be understood from a linear approximation based on the Schr\"odinger equation. The solution to the Schr\"odinger equation for a single particle with energy $E$ impacting a square potential barrier as in Eq. (\ref{barrier}) is known, and the transmission coefficient is given by
\begin{gather}
\begin{aligned}
T=&\Theta (E)\times \\
&\begin{cases}
\Bigl[1 + \frac{V_0^2}{4E(V_0-E)}\sinh^2 \left(b\sqrt{2(V_0-E)}\right)\Bigr]^{-1} \\
\phantom{{}\Bigl[1 + \frac{V_0^2}{4E(V_0-E)}\sinh^2 \left(b\sqrt{2()}\right)} \text{for $E<V_0$}\\
\Bigl[1 + \frac{V_0^2}{4E(E-V_0)}\sinh^2 \left(b\sqrt{2(E-V_0)}\right)\Bigr]^{-1} \\
\phantom{{}\Bigl[1 + \frac{V_0^2}{4E(E-V_0)}\sinh^2 \left(b\sqrt{2()}\right)} \text{for $E>V_0$},
\end{cases}
\end{aligned}
\label{Transmission}
\end{gather}
where the Heaviside step function, $\Theta (E)$, expresses that the energy of the particle scattered must be positive, which is evident in the single particle case, but will be of importance in the following. Ignoring the non-linear interaction term in the GPE is not valid in regions of space where the strong potential is not present. In these regions the soliton behavior is well known, and in particular we know the average energy per particle $\epsilon$:
\begin{equation}
\epsilon =\frac{E_\mathrm{B}}{N}=\frac{1}{2}v^2-\frac{N^2}{24}.
\label{epsilonsol}
\end{equation}
Because of the attractive interactions between the particles, $\epsilon$ becomes negative when $v < \frac{N}{\sqrt{12}}$ (we have chosen $N=1$ in our calculations which gives $\frac{N}{\sqrt{12}}\approx 0.29$ in our dimensionless units).\par
To describe scattering of the soliton in the linear approximation, it makes good sense to replace the energy $E$ in Eq. (\ref{Transmission}) by the average energy per particle in the soliton, $\epsilon$ in Eq. (\ref{epsilonsol}), and in the velocity regime $v \leq \frac{N}{\sqrt{12}}$ the Heaviside stepfunction in Eq. (\ref{Transmission}) suppresses the transmission and predicts full reflection of the soliton. The prediction based on this description is shown in Fig. \ref{kvant} and is in very good agreement with the solution of the GPE. The small transmission probability observed below the velocity threshold may be due, in parts, to our use of wave packets with finite spread, and hence with a momentum distribution exceeding the threshold value.

\begin{figure}[htbp]
\myfigfont
\centering
\begin{minipage}[t]{0.47\textwidth}
\includegraphics[scale=0.96]{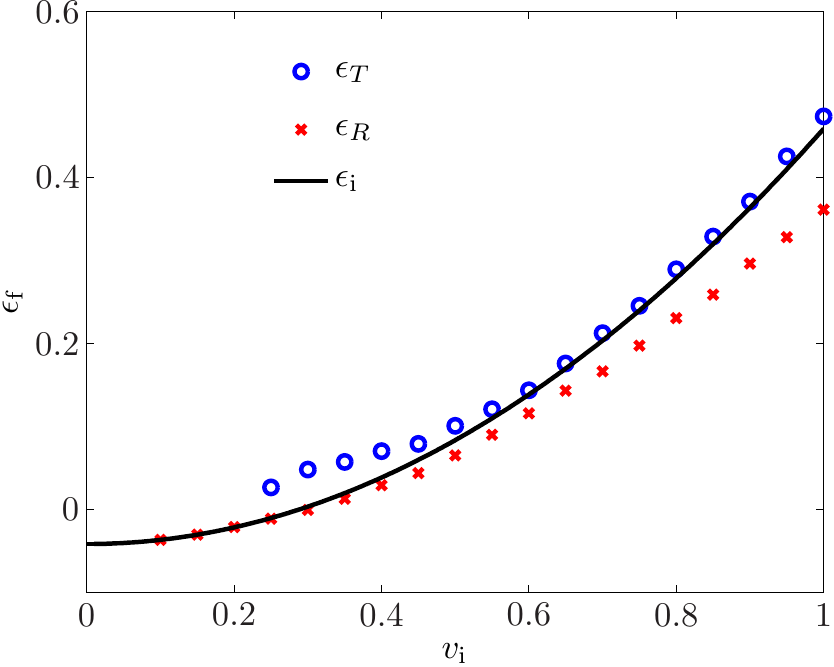}
\end{minipage}
\begin{minipage}[b]{0.47\textwidth}
\caption{(Color online) The energy per particle as a function of the incoming soliton velocity. After the scattering the energy per particle is found by numerical simulations of the GPE and Eq. (\ref{epsiloniandf}) (symbols) and the energy per particle of the incoming soliton (solid line) is found by Eq. (\ref{epsilonsol}). }
\label{Efinal}
\end{minipage}
\end{figure}

We expect the energy per particle of the reflected and of the transmitted part of the soliton to equal the energy per particle of the initial soliton, $\epsilon_\mathrm{i}$. Since the transmitted (reflected) parts contains a fraction $N_T=T\cdot N$ ($N_R=R\cdot N$) of the atoms, the interaction energy per particle acquires a new value, and we thus expect the velocity of the components to change according to the equation
\begin{equation}
\epsilon_\mathrm{i} = \frac{1}{2}v_\mathrm{i}^2-\frac{N^2}{24}=\frac{1}{2}v_{T/R}^2-\frac{N^2_{T/R}}{24}.
\label{epsiloniandf}
\end{equation}
Fig. \ref{Efinal}  shows examples of the energy per particle in the transmitted and reflected components, determined by the GPE and by the expression (\ref{epsiloniandf}). There is good agreement between the energy per particle in the incoming and in the strongest of the outgoing components, i.e., with the reflected (transmitted) component for low (high) velocities. The poor agreement for the weakest outgoing component is a signature that it is not a pure soliton solution. 

\section{Scattering of Dark-Bright Solitons}
\label{Scatteringofdarkbrightsolitons}
Dark and dark-bright solitons retain their localized nature when they collide or propagate in slowly varying potentials \cite{bog,10,8,Becker,BogA,Middelkamp,mangevector,darkbrightN}.\par
We will here study collisions of dark-bright vector solitons on the rapidly varying square potential, which we choose to only directly affect the bright component. The coupled GP equations for this problem are then given by
\begin{gather}
\begin{aligned}
&i\frac{\partial \psi_\mathrm{D}(x,t)}{\partial t}=\Biggl(-\frac{1}{2}\frac{\partial ^2}{\partial x^2}+\abs{\psi_\mathrm{D}(x,t)} ^2\\
&\phantom{{}i\frac{\partial \psi_\mathrm{D}(x,t)}{\partial t}} +\abs{\psi_\mathrm{B}(x,t)} ^2\Biggr)\psi_\mathrm{D}(x,t),\\
&i\frac{\partial \psi_\mathrm{B}(x,t)}{\partial t}=\Biggl(-\frac{1}{2}\frac{\partial ^2}{\partial x^2}+\abs{\psi_\mathrm{B}(x,t)} ^2\\ 
&\phantom{{}i\frac{\partial \psi_\mathrm{B}(x,t)}{\partial t}} +\abs{\psi_\mathrm{D}(x,t)} ^2 + V(x)\Biggr)\psi_\mathrm{B}(x,t)\psi_\mathrm{B}(x,t).
\end{aligned}
\label{twocomponentBECwithV}
\end{gather}
Because of the rapid spatial variation at the edge of the square potential, we expect the vector soliton to break up into a reflected and a transmitted part as we observed for the single bright soliton. Due to the interaction with the dark component, this system is more complicated, and the linear approximation successfully applied to the bright soliton is not valid here.
\begin{figure*}
\myfigfont
\centering
\subfloat[\label{bright}]{\includegraphics[scale=1]{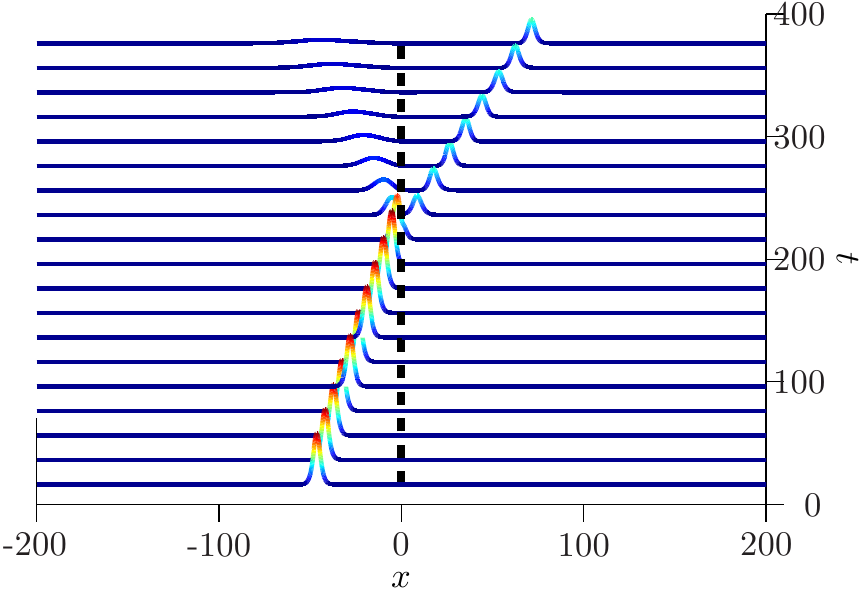}}
%\qquad
\subfloat[\label{dark2}]{\includegraphics[scale=1]{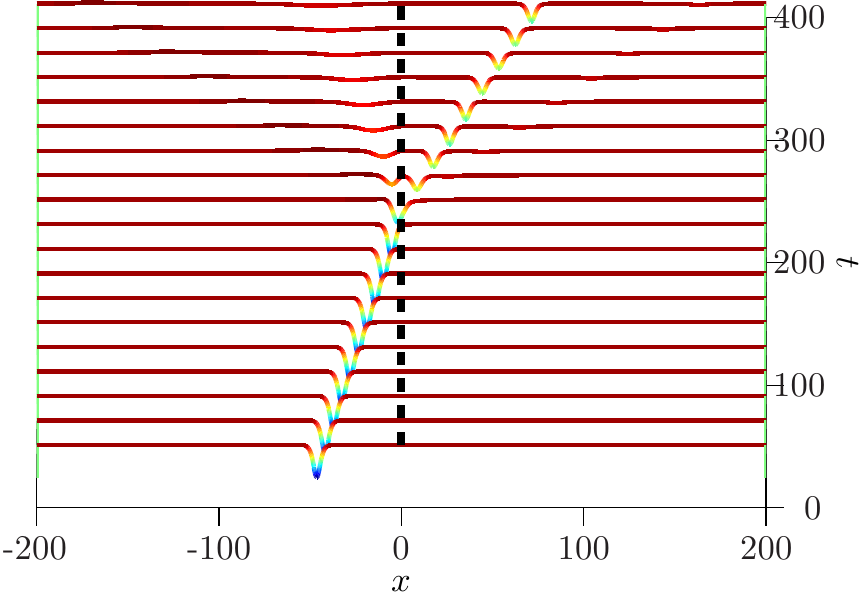}}
\caption{(Color online) Time evolution of the condensate density when the dark-bright soliton scatters on a barrier. \protect\subref{bright} shows the bright component and \protect\subref{dark2} the dark component of the condensate. The incoming soliton has $N_\mathrm{B}=3$ and $\sin\alpha=0.5$, and the barrier, which only directly affects the bright component, has height $V_0=0.2$ and width $b=2$, its location is indicated by the dashed lines. }
\label{timedensity}
\end{figure*}

The results, shown in Fig. \ref{timedensity}, are obtained by numerical simulations of the system (in all numerical calculations $\mu = 1$). The transmission and reflection coefficients, $T_\mathrm{B}$ and $R_\mathrm{B}$, are defined in the same way as for the single bright soliton in Eq. (\ref{TandRnum}). For the dark component, we define the reflection and transmission coefficients as the number of atoms missing with respect to the homogeneous background condensate, on the reflection and transmission side of the barrier, divided by the missing number of atoms $N_D$ in the incident dark component of the vector soliton,
\begin{gather}
\begin{aligned}
&T_\mathrm{D}=\frac{1}{N_\mathrm{D}}\int _0^{x_{\mathrm{sw},T}} \intpar{x} \left(\mu-\abs{ \psi_\mathrm{D}} ^2\right),\\
&R_\mathrm{D}=\frac{1}{N_\mathrm{D}}\int_{x_{\mathrm{sw},R}}^0 \intpar{x} \left(\mu-\abs{ \psi_\mathrm{D}} ^2\right).
\end{aligned}
\end{gather}
The limits $x_{\mathrm{sw},T}$ and $x_{\mathrm{sw},R}$ are determined such that the amplitude variations due to sound waves emitted into the $\psi_\mathrm{D}$-component during the scattering process do not contribute.
\begin{figure*}
\myfigfont
\centering
\subfloat[\label{NB3v05}]{\includegraphics[scale=1]{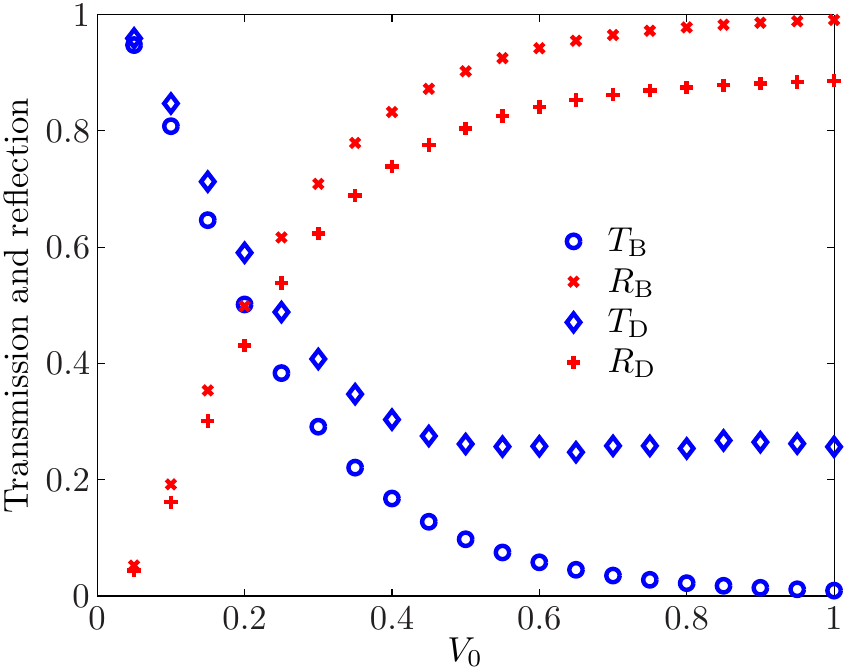}}
\qquad
\subfloat[\label{NB1v05}]{\includegraphics[scale=1]{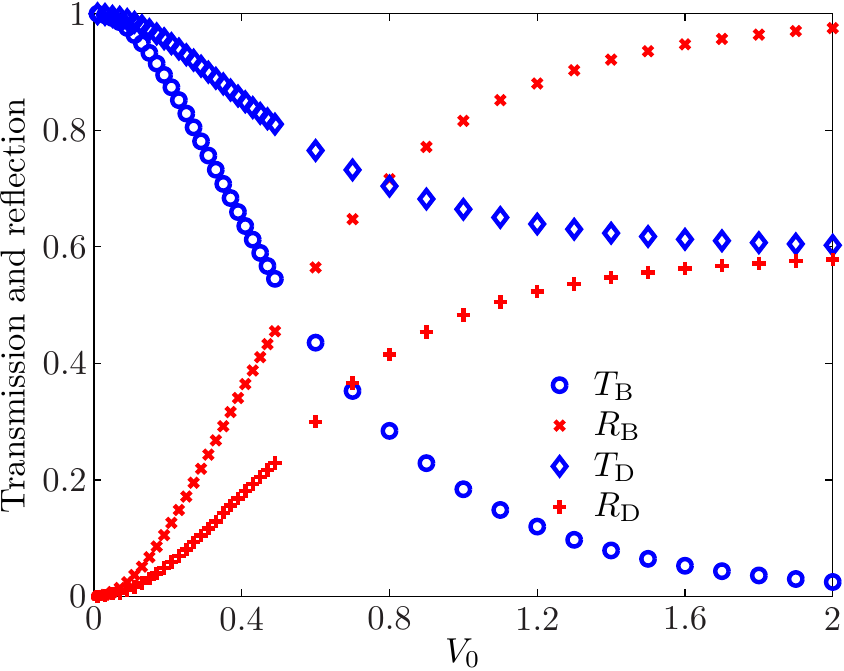}}
\caption{Transmission and reflection coefficients of the dark soliton and the bright soliton as a function of the barrier height $V_0$. In \protect\subref{NB3v05} $N_\mathrm{B}=3$ and in \protect\subref{NB1v05} $N_\mathrm{B}=1$. In both cases $\sin\alpha=0.5$ and the barrier width is that of a dark-bright soliton with $\sin\alpha =0$ ($\kappa ^{-1}$, Eq. (\ref{kappaligning})), for $N_\mathrm{B}=3$ that is $b=2$ and for $N_\mathrm{B}=1$ it is $b\approx 1.3$.}
\label{TRv05}
\end{figure*}
We note from Fig. \ref{TRv05} that both the bright and the dark soliton are indeed split by the potential even though only the bright component is directly affected by the barrier. The dark soliton, however, is not reflected to the same degree as the bright soliton, and when the bright soliton reflection reaches unity, the dark soliton transmission converges to a constant which is independent of a further increase in the barrier height, see Fig. \ref{TRv05}. This is also the case for other properties of the system such as the velocities of the outgoing solitons and the amplitudes of the sound waves emitted in the dark component.

In Fig. \ref{TRv05} it is also seen that the reflection coefficient and the transmission coefficient of the dark soliton do not add up to one when the barrier height is increased. This we ascribe to sound waves carrying a corresponding surplus of atoms in the background condensate. 

When $N_\mathrm{B}$ is changed, it alters the soliton scattering properties since the coupling to the bright soliton, mediating the effects of the barrier, is stronger when $N_\mathrm{B}$ is larger.

\section{Conclusion}

We have investigated scattering of solitons in Bose-Einstein condensates on localized potentials. The homogeneous Gross-Pitaevskii equation supports matter wave solitons, and we have focused on two types of solitons: Bright solitons, which are wave packets without dispersion due to the attractive, non-linear interactions between the atoms in the soliton, and dark-bright solitons in two separate condensate components. In the latter the bright soliton is stabilized by the presence of the dark soliton, a dip in the condensate density. 

In slowly varying potentials, solitons are known to stay intact and their center of mass motion can be described by a classical equation of motion. We have here considered scattering of solitons on square barriers which vary rapidly at the edges, and where a more quantum mechanical behavior of the soliton is observed.

For the bright soliton we found that the scattering properties of the soliton on a localized potential resembles that of classical or quantum mechanical scattering depending on the height and width of the barrier.

When the barrier is low compared to the chemical potential of the soliton, the soliton behaves like a robust classical particle, since the barrier cannot overcome the internal interactions of the soliton. The potential appearing in the resulting classical equation of motion describing the soliton center of mass motion is an effective potential modified by the non-linearity.

When the barrier is high compared to the chemical potential of the bright soliton, the soliton behaves like a single quantum mechanical particle with a modified relation between its energy and velocity. In this regime, the internal interactions are no longer able to keep the soliton intact and it is split into a reflected and a transmitted part. 

The scattering properties of a dark-bright soliton on a barrier only affecting the bright component could be expected to resemble that of a single bright soliton scattering on a barrier. But we find that the scattering of a dark-bright soliton is in fact very different. The differences is due to the interaction between the two components of the dark-bright condensate, which enables an indirect interaction between the dark soliton and the external potential. This interaction is strongest when the bright component is large compared to the dark one. 

Experimental realization of solitons in 1D-condensates is well developed \cite{exppot}, and dark-bright solitons have previously been studied in slowly varying potentials. Thus the prospects of experimental studies along the lines of our calculations seem quite feasible.\par
\FloatBarrier

\bibliography{bibliografi}
\end{document}